\documentclass{article}

\usepackage{bm}

\title{Limits on charge non-conservation from possible seasonal
variations of the solar neutrino experiments} 
\author{
Manuel Torres\thanks{e-mail: torres@fisica.unam.mx} and
Hector Vucetich \thanks{Corresponding author. e-mail:
  vucetich@fisica.unam.mx; Fax: (52) 56225015. 
 On leave of absence
  from: Observatorio 
  Astron\'omico, Universidad Nacional de La Plata.}  \\
Instituto de F\'{\i}sica, 
   Universidad Nacional Aut\'onoma de M\'exico\\ Aptdo. Postal
   20-364, M\'exico D.F. 01000, M\'exico.
       }

\usepackage[numbers]{natbib}
\begin{document}

\maketitle
\begin{abstract}
 Variable speed of light (VSL) theories generically lead to large
violations of charge conservation that can be written in terms of a
dimensionless parameter $\lambda$.  It is shown that the motion of the
Earth with respect to the Sun could lead to a seasonal variation for
the SAGE and GALLEX-GNO experiments and analyzing the reported
counting rates for these experiments, a very stringent bound $\lambda
\le 2 \times 10^{-19} $ is obtained, some $10^9$ times smaller than
previous ones.  Furthermore, a bound on the lifetime of the
$\null^{71}{\rm Ga} \to \null^{71} {\rm Ge}$ charge-nonconserving
decay in VSL theories is found as: $ \tau_{\rm CNC} \ge 1.4 \times
10^{27} \,\, {\rm years}$. Similarly a new upper limit for the ratio
of the charge-nonconserving to the normal weak decay of the neutron in
VSL theories is obtained: $\Gamma(n \to p + \nu_e + \bar{\nu}_e)
/\Gamma(n \to p + e + \bar{\nu}_e) \le 2 \times 10^{-27}$.
\end{abstract} 

\begin{description}
\item[PACS:] 11.30.-j,04.50.+h,26.65.+t

\item[Keywords:] Charge conservation; VSL Theories; solar neutrino
experiments. 
\end{description}

Recently, a growing interest in the time variation of fundamental
constants has arisen, driven by the tantalizing observation that
$\alpha = e^2/\hbar c$ may have been slightly smaller when the
Universe was younger $z > 1, t_0 - t > 8 \times 10^9 \rm{yr}$
\cite{Webb:1998cq,Webb:2000mn}. This may be attributed either to a
time variation of the electron charge $e$ or of the speed of light
$c$ \cite{MBS02,Barrow:2002hi,Peres:2002ki}.

The second possibility has received much attention, since a varying
speed of light may solve the homogeneity, horizon and flatness
problems in cosmology
\cite{Albrecht99,Barrow:1998ef,Barrow:1999jq}. Several variations of
these theories have been proposed with varying degree of
sophistication
\cite{Moffat93a,Moffat93b,Barrow99,Magueijo00a,Magueijo00b}.

It was observed in Ref. \cite{Landau:2001qb} that a generic variable
speed of light (VSL) theory leads necessarily to a large violation of
charge conservation. This is because Maxwell's equations are modified
in the form:
\begin{equation}
  \frac{1}{c} \partial_\mu\left(cF^{\mu\nu}\right) = 4\pi j^\nu,
					\label{ModMaxwell}
\end{equation}
from which the following relation of charge
nonconservation to the variation of $c$ can be derived
\cite{Landau:2001gz} 
\begin{equation}
 \frac{\dot Q}{ Q }   =   -  \frac{\dot c}{ c }
  \label{qc}\, ,
\end{equation}
This equation provides very stringent tests of the variation of $c$,
since there have been many experiments to test the conservation of
charge \cite{Landau:2001gz}.

If it is assumed that the electron charge $e$ is constant, charge
non-conservation can only be the result of processes that change
charge discontinuously. Models for these kind of processes have been
proposed in Refs. \cite{Magueijo00a,harko}.  An analysis based on this
models yield an upper limit on ${\dot c } /c$ that are much more
smaller than those obtained by direct measurements \cite{Prestage95}:
$\vert {\dot c } /c \vert < 10^{-31} \,\, yr^{-1}$.

A consistent VSL leads to a dynamically prediction of the value of $c$
via a wave-like equation for $\psi = \ln (c/c_0)$, whose source term is
proportional to the trace of the energy-momentum tensor. Hence, in the
neighborhood of a quasistatic system, such as a star, a generic
expression for the variation of $c$ takes the form
\cite{Magueijo00b,Landau:2001gz} 
\begin{equation}
c(t) \, = \, c_c(t) \, \left ( 1 - { \lambda \frac{G M}{c_0^2 r} }\right)
 \label{varc1}\, .
\end{equation}
Here $\lambda$ is a constant that depends on the specific VSL theory. 
In the limit $r \to \infty$ the expression for $c$ reduces to the
cosmological one $(c \to c_c(t))$. In this work we are interested in
finding a bound to the constant $\lambda$ that parametrizes the charge
nonconservation.

In the static regime, the $\bm{\nabla} c$ contributions are second
order and can be neglected. In this case there are two contributions
to the charge non-conservation: one accounts for time-variation over
cosmological time scales and the other for the motion of the earth
with respect to massive bodies 
\begin{equation}
\frac{\dot c}{ c }  =  \left( \frac{\dot c}{ c }
 \right)_{\rm cosmological}  +  \left( {{\dot c} \over c }
 \right)_{\rm local}  =  n \frac{\dot a}{ a } + \frac{ \lambda G M}{
 c_0^2 r ^3 } \, \bm{r} \cdot \bm{ v }
 \label{varc2} \, ,
\end{equation}
where $\bm{ v} $ is the velocity of the Earth with respect to the
center of mass of nearby bodies. 
The last term of equation (\ref{varc2}) represents  a breakdown
of charge conservation, induced by the presence of local matter.

Also, charge nonconservation may be associated to the breakdown of
Einstein's Equivalence principle, as shown in
Ref. \cite{Landau:2001qb}. In this case, CNC processes would lead to
an expression similar to the second term of (\ref{varc2}) 
\begin{equation}
\frac{\dot{Q}}{Q} = \frac{\Gamma_0}{4} \frac{{\bm{g}}_0 \cdot
{\bm{v}}}{c_0^2} \label{Var-N-Def}
\end{equation}
where $\bm{g}_0$ is the local gravitational acceleration and the
parameter $\Gamma_0$ characterizes anomalous accelerations and
anomalous mass tensors \cite{WillTh}:
\begin{eqnarray}
\delta m_P &=& 2 \Gamma_0 \frac{E_C}{c_0^2}, 
\label{Def-delta-m}\\
\Delta a_C &=& \frac{\delta m_P}{m} g_0. 
\label{Def-Delta-a}
\end{eqnarray}

The SAGE
\cite{Abdurashitov:1999bv,Abdurashitov:1999zd,Abdurashitov:2002xa} and
GALLEX-GNO 
\cite{GALLEX-I,GALLEX-II,GALLEX-III,GALLEX-IV,Altmann:2000ft,%
Cattadori:2002rd} radiochemical solar neutrino experiments are based
on the neutrino capture reaction $\nu_e + \null^{71}{\rm Ga} \to e^- +
\null^{71}{\rm Ge}$. This reaction has a low energy threshold: $E_{th}
= 0.2332 \, {\rm MeV}$ that makes possible the detection of $pp$ solar
neutrinos. However it was pointed out in Ref. \cite{Norman:1996eg} that in
these experiments the appearance of $ \null^{71} {\rm Ge} $ produced
by the $\nu_e$ on the target $\null^{71} {\rm Ga}$ can in principle be
the result of charge nonconservation reaction. The reason is that as
there is no electron emitted in a charge nonconservation reaction such
as $ n \to p + \gamma$ or $n \to p + \nu_ e + {\bar \nu}_e$, this
leads in principle to the possibility that the $ \null^{71} {\rm Ga}$
undergoes a charge-nonconserving decay to the ground and to the first
excited state of the $ \null^{71}{\rm Ge} $, that otherwise wold be
forbidden by energy conservation.  Based on the results
form SAGE and GALLEX alone it is not possible to discern whether the $
\null^{71}{\rm Ge} $ was produce via a charge-nonconserving processes
or by a normal weak decay; hence it is possible to test those theories
that predict the violation of the charge conservation.

In Ref. \cite{Landau:2001gz} these results were applied for the case
of the Earth motion relative to Virgo supercluster, in this case $ v
\approx 1000 \, {\rm km/s} $ and $G M / c_0^2 r \approx 2 \times
10^{-7}$. The charge nonconservation effect would contribute to the
CNC $\null^{71} {\rm Ga} \to \null^{71}{\rm Ge} + {\rm
neutrals}$ processes, hence an upper limit on the parameter $\lambda $
can be derived: 
\begin{equation}  
\vert \lambda \vert \le 2 \times 10^{-10}. \label{PrevBound}
\end{equation}

The above limit is based on cosmological
considerations. Ref. \cite{Magueijo00b} has stressed the importance of
local tests of variable speed of light theories. In this paper, we
shall develop such a test, based on changes in the rate of Charge
Nonconserving Processes (CNC) induced the motion of the Earth around
the Sun, that introduces a periodic
variation of $c$ with a one year period \cite{RS1}:
\begin{equation}
\frac{\dot{Q}}{Q} \simeq -\lambda  \frac{G M_\odot}{c^2_0 a_\oplus}
	e_\oplus \omega_\oplus \sin\omega_\oplus(t - t_0)
					\label{Qloc-Earth} 
\end{equation}
where $M_\odot$ is the mass of the Sun, $ a_\oplus$ and $e_\oplus
\approx 0.0167$ are
the major axis and eccentricity of the orbit of the Earth, and $
\omega_\oplus = 2\pi/T_\oplus$ the frequency (mean motion) of the
Earth motion around the Sun. Since $t_0$, the time of perihelion
passage, is around January 10, a periodic signal may be discerned (or
bounded) comparing spring and autumn results from experiments testing
charge conservation.  In the case of the Gallium experiments a seasonal
effect is expected; the term $\bm{ r \cdot v}$ varies along the year
in a sinusoidal way.  As previously discussed, a charge
nonconservation decay $\null^{71} {\rm Ga}$ to $\null^{71} {\rm Ge}$
is energetically allowed.  According to Eqs. (\ref{qc}) and
(\ref{varc2}) a contribution to the charge increase required in this
situation can only arise during half of the year, because the term
$\bm {r \cdot v}$ changes sign during the other half and the reverse
process ($\null^{71} {\rm Ge}$ to $\null^{71} {\rm Ga}$) is
energetically forbidden. Hence the
average contribution of the charge nonconserving effect to the
$\null^{71} {\rm Ge}$ counting for the spring period would be
expected to be different as compared to that of autumn period.

Combining the relevant terms of Eqs.(\ref{qc}) and (\ref{varc2}) the
relative charge variation can be written as
\begin{equation}
 \frac{\dot Q}{ Q }  =  \frac{ \lambda G M }{ c_0^2 \, a_\oplus }
  \Lambda  \label{qc2}\, ,
\end{equation}
Here $G M / c_0^2 \, a_\oplus \approx 1 \times 10^{-8}$, and $\Lambda $ is
the time average of the $\bm r \cdot \bm v /r^2$ factor, that for the
spring period can be computed as
\begin{equation}
 \Lambda = \frac{2 e_\oplus h_\oplus}{ \left( 1 + e_\oplus \right)^3
 a_\oplus^2} \left(  1 + \frac{ e_\oplus^2}{ 3} \right) 
  \label{lambm}\, ,
\end{equation}
where  $ h = L /m $ is the angular
momentum per unit mass of the Earth and the rest of the notation is
the same as in Eq. (\ref{Qloc-Earth}). Hence $h_\oplus /a_\oplus \approx
29.79 \, {\rm km /s}$ corresponds to the mean orbital velocity.

The SAGE and GALLEX collaborations have reported results of $73 \pm
18$ and $78 \pm 12$ solar neutrino units ($\rm SNU$) respectively,
where $1$ ${\rm SNU}=10^{-36}$ neutrino captures per target atom per
second. Furthermore the SAGE collaboration have reported the neutrino
rates for periods of 30 days along the whole year. Unfortunately, the
GALLEX-GNO collaboration have not reported monthly or bimonthly
averages. 
The seasonal averages where computed from the data for individual runs in
Refs. \cite{GALLEX-I,GALLEX-II,GALLEX-III,GALLEX-IV,Altmann:2000ft,%
Cattadori:2002rd} and from the monthly data in
Refs.
\cite{Abdurashitov:1999bv,Abdurashitov:1999zd,Abdurashitov:2002xa}.
Analyzing these 
results the asymmetry for the spring period as compared with
autumn one yields 
\begin{eqnarray}
  \Delta\Phi_{\rm GALLEX+GNO} &=& 11.3 \pm 9.8\; {\rm SNU}
      \nonumber\\
  \Delta\Phi_{\rm SAGE}       &=& -0.5 \pm 10.5\; {\rm SNU}
       \nonumber\\
  \Delta\Phi_{\rm Comb}       &=& 5.8  \pm 7.2\; {\rm SNU}
        \label{S-A-Asym}
\end{eqnarray}

 The maximum
possible flux asymmetry $\Delta \Phi$ when these two periods are
compared can be taken as a $3\sigma$ limit:
\begin{equation}
  \mid \Delta \Phi\mid \leq 21.6\; {\rm SNU}. \label{AsymUB}
\end{equation}
  From this result we conclude that the charge asymmetry ${\dot Q / Q
}$ can be no larger than $21.6 \times 10^{-36} \, {\rm s^{-1}}$ per
target atom; in combination with the previous results it leads to the
following bound for the parameter $\lambda $:
\begin{equation}
\mid \lambda \mid \le 2 \times 10^{-19}  \,. \label{bl}
\end{equation}
This bound is about $10^{9}$ times smaller than the previous one,
based on cosmological considerations.

A similar bound would be found for the breakdown of Einstein's
Equivalence Principle:
\begin{equation}
  \mid \Gamma_0 \mid \leq 10^{-18}.
\end{equation}
This is somewhat smaller than the expected accuracy of STEP
\cite{STEP93,STEP94}. 

These results can also be used to derive a bound for the $\null^{71}
{\rm Ga} \to \null^{71} {\rm Ge} + {\rm neutrals}$ lifetime, in the
case of processes of the type proposed in Refs
\cite{Magueijo00a,harko}. From the limits on the $\Delta \Phi = 21.3
    \; {\rm SNU}$ asymmetry we conclude that the charge-nonconserving
     decay rate of 
$\null^{71} {\rm Ga} \to \null^{71} {\rm Ge} $ fulfills the relation
\begin{equation}
 \frac{d N \left( \null^{71} {\rm Ga} \right)}{ dt }  = 
  \lambda_{\rm CNC} N \left( \null^{71} {\rm Ga} \right) < \Delta \Phi N \left(
  \null^{71} {\rm Ga} \right) 
   \label{life}\, 
\end{equation}
where $N \left( \null^{71} {\rm Ga} \right)$ is the number of $ \null^{71}
{\rm Ga} $ atoms in the neutrino detectors and $ \lambda_{\rm CNC} $ is the
 decay rate for the $\null^{71} {\rm Ga} \to \null^{71} {\rm Ge} $
charge-nonconserving decay.  Hence the bound for the lifetime of this
processes yields
\begin{equation}
 \tau_{\rm CNC}\left(\null^{71} {\rm Ga} \to \null^{71} {\rm Ge}\right)
 \ge  1.4  \times 10^{27}  {\rm years}   \label{life2}\, . 
\end{equation}
This result should be compared with the previous upper limit $
\tau_{\rm CNC} \ge 3.5 \times 10^{26} \, {\rm years} $ obtained in
Ref. \cite{Norman:1996eg}. This bound was obtained by considering the
the total counts of $73 \pm 18$ $SNU$ for SAGE and $78 \pm 12$ $SNU$
for GALLEX.  In the same way, a bound for the branching ratio of CNC
processes to beta decay can be obtained as in
Ref. \cite{Norman:1996eg} from the seasonal asymmetry effect produced
by the second term of Eq. (\ref{varc2}).
\begin{equation}
\Gamma(n \to p + \nu_e + \bar{\nu}_e) /\Gamma(n \to p + e +
\bar{\nu}_e)  \le 2 \times 10^{-27} 
\end{equation}

We have shown that the absence of seasonal spring-autumn variations in
Gallium solar neutrino experiments imposes extremely tight bounds on
the parameters of varying speed of light theories, some $10^9$ times
smaller than those obtained from cosmological considerations. This not
only confirms the importance of local tests of VSL theories, as
stressed in Ref. \cite{Magueijo00b} but also imposes very strict
constraints on their structure.

H. V. acknowledges the hospitality of IFUNAM. This work was partially
supported by CONACyT grants No. \texttt{G32736-E} and \texttt{U42046}.

\bibliography{exgrv,tvref_ph,tvref_th,VSL,neutrinos,RPP}

\end{document}